\documentclass[aps,pre,reprint]{revtex4-2}
\usepackage{amsmath}
\usepackage{amsmath}
\usepackage{amssymb}
\usepackage{graphicx}
\usepackage{color}
\usepackage{subfigure}           
\usepackage{natbib}                                                                
\usepackage[usenames,dvipsnames]{xcolor}
\usepackage{bm}
\usepackage{epstopdf}
\definecolor{brick}{RGB}{150,41,56}

\newcommand{\nhs}[1]{\textcolor{black}{#1}}
\newcommand{\modified}[1]{\textcolor{black}{#1}}
\newcommand{\removedNH}[1]{}

\renewcommand{\vr}{\mathbf{r}}

\newcommand{\equref}[1]{Eq.~(\ref{#1})}
\newcommand{\equrefs}[1]{Eqs.~(\ref{#1})}

\newcommand{\figref}[1]{Fig.~(\ref{#1})}

\begin{document}
\title{Transport and phase separation of active Brownian particles in fluctuating environments}

\author{S. M. J. Khadem$^{1}$, N. H. Siboni$^1$, and S. H. L. Klapp$^1$}
\affiliation{$^1$\mbox{Institut f\"ur Theoretische Physik, Technische Universit\"at Berlin}, \\Hardenbergstra\ss e 36, 10623 Berlin, Germany}
\date{\today}
\begin{abstract}
   In this work, we study the dynamics of a single active Brownian particle, as well as the collective behavior
  of interacting active Brownian particles, in a fluctuating heterogeneous environment. We employ 
  a variant of the diffusing diffusivity model where  the equation
  of motion of the active particle involves a time-dependent 
  motility and diffusivities. Within our model, those fluctuations are coupled to each other. Using analytical methods, 
 we obtain the probability distribution function of  particle displacement and its moments for a single particle. We 
 then investigate the impact of the environmental fluctuations on the collective behavior of the active Brownian 
  particles by means of extensive numerical simulations. Our results show that  the fluctuations hinder the motility-induced phase separation, accompanied by a significant change of the density dependence of particle velocities. These effects are interpreted using our analytical  results for the dynamics of a single particle.
\end{abstract}

\maketitle


\section{Introduction}
\label{intro}

Transport of a tracer particle in  complex heterogeneous environments  such as biological media~\cite{mckinley2009transient, hofling2013anomalous}, porous material~\cite{koch1988anomalous} or polymeric networks~\cite{wong2004anomalous}  can widely  differ from the normal diffusion of a passive particle in a simple fluid. Normal diffusion is characterised by a Gaussian probability distribution of the particle displacement  and  a mean squared displacement~(MSD) which  grows linearly in time, yielding a constant long-time diffusion coefficient. An often observed phenomenon in   complex, crowded environments is  anomalous diffusion of a (passive) tracer particle, described by a nonlinear MSD associated with either a Gaussian or non-Gaussian probability distribution \cite{klages2008anomalous}. To explain the observed anomalies,  a variety of  stochastic models,  mainly based  on   generalization of  Einstein's and Langevin's works on normal diffusion, have been proposed (for a review, see, e.g.,~\cite{sokolov2012models}).

 In many heterogeneous systems, however, a tracer particle  experiences varying diffusivities over time. This is either due to the dynamical  evolution of the  surrounding   medium  by its own, or due to the motion of the particle in regions with different diffusivities.  Such situations have been observed in numerous single particle tracking experiments, for widely different types of (passive) tracer particles and environments~\cite{exp1,exp2,exp3,exp4},  as well as in several simulation studies~\cite{sim1,sim2}. In these studies, it was shown that, while the MSD  of the passive  particle remains linear in time, its displacement probability density function~(PDF) deviates from the Gaussian form at short times. These observed deviations  were in contradiction with Fick's second law, where the displacement PDF    obeys the standard diffusion equation and has a Gaussian form at all times. These observations motivated the development of a new class of  stochastic processes, that is,  'Brownian yet non-Gaussian'~\cite{metzler2017gaussianity}. 
 
 This new dynamics was first theoretically explained by  the concept of  superstatistics~\cite{beck2003superstatistics}, where a single Gaussian distribution is averaged over   an exponential  distribution  of  diffusivities, $p(D)$ \cite{exp3}. Afterwards, another approach termed as   'diffusing diffusivity' was   put forward~\cite{chubynsky2014diffusing, jain2016diffusion, jain2016diffusing} assuming that the diffusivity constantly changes in time according  to a specific stochastic process. The diffusing diffusivity model was  further studied with respect to time-averages and  ergodicity breaking properties \cite{lanoiselee2018model}. It was also compared  with the concept of  the generalized Grey-Brownian motion with random diffusivity~\cite{sposini2018random}. An elegant derivation of the diffusing diffusivity model using the  sub-ordination approach  was later introduced in Ref.~\cite{chechkin2017brownian}, which connected the superstatistics approach with the diffusing diffusivity model. We note that the dynamics of the 'Brownian yet non-Gaussian' process can also be reproduced by the  continuous time random walk model in equilibrium with a truncated, heavy tailed waiting time PDF~\cite{khadem2017nonscaling}.
 
The aforementioned  studies  mainly focus on the transport of passive tracer particles. Here, we are rather interested in active particles~(AP), such as bacteria or self-propelled colloids. 
Indeed, similar to the case of passive particles, active particles  are also often found  in heterogeneous environments.
Examples  include  molecular motors inside a  'crowded' environment of  cells~\cite{R},  bacteria in highly complex tissues~\cite{S} or  herds of animals migrating in forest~\cite{T}. Most of the non-equilibrium features of the collective behavior of APs and their individual dynamics are strongly affected, when they take place in heterogeneous media~\cite{Peruani2015, Peruani2013b, Quint2015}.
Recent studies discussed different phenomena, such as anomalous diffusion of AP in a random Lorentz gas~\cite{zeitz2017active}, trapping of APs  in the presence of obstacle arrays~\cite{Peruani2013}, avalanche dynamics transport of run-and-tumble bacteria~\cite{reichhardt2018avalanche} or   clogging and depinning of ballistic active matter~\cite{reichhardt2018clogging} at the macro-scale. These  phenomena  can vary depending  on the type of  APs  as well as on the characteristics of the heterogeneity in the environment, see  Ref.~\cite{bechinger2016active} for a review. From a theoretical point of view, it is therefore quite relevant  to include the environmental heterogeneities in modelling the collective behavior of APs as well as their individual dynamics.

Inspired  by the aforementioned  theoretical works which are restricted to  systems with passive particles,  we  investigate dynamics of APs in temporarily evolving heterogeneous systems. In particular, we explore the impact of fluctuations on  the transport properties of an   active Brownian particle~(ABP)~\cite{ten2011brownian,G}, which is often considered as a minimal model describing   many  experimentally observed dynamics of self-propelled particles~\cite{G, kummel2013circular, kurzthaler2018probing} and the collective behavior of  active colloids and bacteria~\cite{bechinger2016active}. In the first part of the paper, we consider non-interacting  ABPs in a fluctuating medium, focusing on  the behavior of  the MSD and the  PDF of  particle displacement as compared to  the dynamics of ABPs in a homogeneous medium. For those quantities  we calculate general formulas by assuming  that  fluctuations are encoded in a random process with certain properties.  We  then specialize on  square Ornstein-Uhlenbeck process to represent those fluctuations and calculate explicit results for the MSD and the PDF of the particle displacement.  

Some related  studies on APs  in the   presence of  fluctuations can be found in Refs.~\cite{23inlutz, thiffeault2021shake, romanczuk2011brownian}. These articles study the   dynamics of  APs by considering that the activity,  rotational diffusivity, or both, experience  particular forms of   fluctuations. However, fluctuations in those quantities are assumed to be independent from each other.   In the present paper we consider a different situation, where fluctuations of the motility and  diffusivities  are correlated. Such a situation can occur when  the origin of the  fluctuations is external, in the sense that they   arise from the surrounding  environment. Thus, they  affect the activity and diffusivities of the ABP simultaneously and  in the same way. Therefore, our results vastly differ from those in~\cite{23inlutz, thiffeault2021shake, romanczuk2011brownian}.

In the second part of the paper,  we proceed  by studying  the effect of  fluctuations on  the collective behavior of  interacting  ABPs. We focus in particular,  on  the phenomenon of the motility-induced phase separation (MIPS). From  previous studies (see, e.g.,~\cite{cates2015motility}),  it is known that the rotational diffusion and the  motility play essential  roles in the emergence of MIPS. Introducing  fluctuations of these parameters, one therefore expects   a considerable impact on the occurrence of MIPS. Here we explore these  modifications by obtaining the non-equilibrium phase diagram and the connection between the particle velocity and the local density.  To indicate the relevance of our findings, we support our results by calculating  an effective free energy of the system  as a function of local density. 

  To the best of our knowledge, a study of the collective behavior of ABPs with (correlated) fluctuations of motility and diffusivity, has not been done so far. To some extent the present model, where the fluctuations are assumed to be externally induced, could be considered  as a minimal model to describe the collective bahavior of APs in disordered  media~\cite{ro2021disorder}. In fact,  as we will show in the second part of the paper, our external fluctuations have a similar effect as disordered media, that is, a hindrance of MIPS.

\bigskip
\section{Active Brownian particle with fluctuations}
\label{s1}
In this section, we start with a brief  review of the ABP model in homogeneous environment (which we refer to as the conventional ABP model). 
We continue by building up a modified   ABP  model by introducing fluctuations into the conventional
description.  We then analytically investigate the effect of these fluctuations on the PDF of the displacements and the MSD, using the sub-ordination approach. To complete the discussion of the single-particle dynamics, we support our theoretical predictions by numerical results.

\subsection{Conventional Active Brownian Particle}
In the ABP model,  
each (spherical)  particle displays  directed motion in addition to the Brownian motion. 
 This directed motion arises from a self-propulsion force, $F_0$, which is (typically) directed 
  along an anisotropy  axis of the particle, i.e.,  the heading vector.
 Due to  thermal fluctuations, the particle experiences  translational  ($\bm{\xi}_{T}(t)$) and rotational ($\xi_R(t)$) noises, which affect 
 its position (\textbf{r}) and orientation ({$\phi$}), respectively. 	
 The two-dimensional  ABP model is thus be described by the Langevin equations \cite{K}

 \begin{subequations}
  \begin{align}
    \gamma_T \dot{\textbf{r}}(t)&=F_0 \mathbf{\hat{e}}(t) + \sqrt{2 k_B T_T \gamma_T} \bm{\xi}_{T}(t)~  \\
\gamma_R \dot{\phi}(t)&=\sqrt{2 k_B T_R \gamma_R}{\xi}_R(t)~,
  \end{align}
  \label{1}
\end{subequations}

where the vector $\mathbf{\hat{e}}(t):=(\cos\phi, \sin\phi)$ determines the heading direction of the particle within the plane. The 
parameters \mbox{$\gamma_T$} and $\gamma_R$ denote the 
translation and  rotational Stokes friction coefficients, respectively \cite{bechinger2016active}, that is,
 \begin{subequations}
  \begin{align}
\gamma_T&=6 \pi \eta R, & \\
 \gamma_R&= 8 \pi \eta R^3, 
\label{constant-fric}
  \end{align}
\end{subequations}

 with $\eta$ being the 
viscosity and $R$ the radius of the particle. 
The noise terms are Gaussian white noises satisfying  \mbox{$\langle \bm{\xi}_T(t) \rangle=0$}, $\langle \xi_R(t) \rangle=0$, 
\mbox{$\langle{\xi}_{T,j}(t_1) {\xi}_{T,i}(t_2)  \rangle= \delta(t_1-t_2)\delta_{i,j}$} (with $i,j$ being  Cartesian components of $ \bm{\xi}_T(t)$), and $\langle \xi_R(t_1)
\xi_R(t_2)\rangle= \delta(t_1-t_2)$. The parameters 
$k_{B}T$ and $k_{B} T_R$ are the  thermal energies  quantifying the strength of the translational and the rotational 
noises, respectively \cite{lowen2019inertial}. In many applications~\cite{ga, gb,gc,gd},
it is assumed that $T=T_R$. The rotational and  translational diffusion coefficients  then follow as 
\begin{equation}
D_{T(R)}=\frac{k_BT}{\gamma_{T(R)}}.
\label{constant-diff}
\end{equation} 

 Further, the  motility of the ABP is defined as 
\begin{equation}
v_0= F_0/ \gamma_T=F_0/(6\pi\eta R).
\label{constant-active}
\end{equation} 
 As it is implied by Eqs.~(\ref{constant-fric}),~(\ref{constant-diff}) and (\ref{constant-active}), both diffusivities and the motility are  proportional to $\eta^{-1}$ and constant in time.

 \subsection{Modelling active Brownian particle with fluctuations }
   We now relax  the conditions of constant translational and rotational diffusivities.  Specially, we consider a situation where the environment of the particle causes fluctuations affecting both, the motility and the diffusivities. 

 As a justification, we consider a 
  heterogeneous environment whose viscosity  varies in space, that is,  $\eta=\eta(\textbf{r})$. In such an environment, the diffusing particle experiences
  different viscosities while discovering the  medium.  To 'sample' the  heterogeneity, the particle  needs a certain time. For  measurements with
  time laps longer than this time, recovers the conventional ABP model with averaged, i.e.  constant, parameters. However, when the measurement time interval is shorter, the spatial
variation of $\eta$ needs to be taken into account in the equation of the motion of the particle. 


   One approach to describe such a situation is diffusing diffusivity model~\cite{chubynsky2014diffusing, jain2016diffusion, jain2016diffusing}, that is, the spatial heterogeneity of an evolving 
 medium  is  \textit{mimicked} by a  time  dependent stochastic process.   In our case, the characteristic parameter is the viscosity, $\eta(\textbf{r})$,  which we 
assume  to be described by  a random process, 
  $\varUpsilon^{-1}(t)$ (where  $\varUpsilon(t)$ is constrained to be  positive). This implies (see Eqs.~(\ref{constant-fric})) that the friction coefficients become time-dependent quantities, that are proportional to this process, i.e., 
 $\gamma_{T,R}(t) \propto \varUpsilon(t)^{-1}$. 
 Rewriting the Langevin equation in the following form 
 \begin{subequations}
  \begin{align}
 \dot{\textbf{r}}(t)&=\frac{F_0}{\gamma_T(t)} \mathbf{\hat{e}}(t) +\sqrt{ \frac{2k_B T}{\gamma_T(t)}}\bm{\xi}_{T}(t)  \\
\dot{\phi}(t)&= \sqrt{ \frac{2k_B T}{\gamma_R(t)}}{\xi}_R(t)~, 
\end{align}
\label{general-LE}
\end{subequations}
 it becomes evident that the environmental fluctuations affect the motility and diffusivities, given by the time-dependent analogs of Eqs.~(\ref{constant-active}) and (\ref{constant-diff}), simultaneously and in the same way. As a consequence  one has

 \begin{eqnarray}\label{eq:assumption1}
 D_R(t)/D^{\star}_R&=& D_T(t)/ D^{\star}_T= v_0(t)/v^{\star} = {\varUpsilon(t)},
\label{sync}
\end{eqnarray}
where the quantities with star denote reference values. 

 Combining Eqs.~(\ref{general-LE}) and (\ref{sync}), we obtain the following modified Langevin 
 equation of the ABP
\begin{subequations}
  \begin{align}
\dot{\textbf{r}}(t)&=v^{\star} \varUpsilon(t) \mathbf{\hat{e}}(t) +\sqrt{2D^{\star}_T\varUpsilon(t)} \bm{\xi}_{T}(t), \\
\dot{\phi}(t)&=\sqrt{2D^{\star}_R\varUpsilon(t)} \xi_R(t).
\end{align}
\label{mle}
\end{subequations}
 
 These equations have to be supplemented by an equation determining the time-dependence of the
 process $\varUpsilon(t)$. This will be done at a later stage, see~Section~(\ref{Theory-results}). 
A direct consequence of  Eq.~(\ref{sync}) is that, despite the time dependency of the motility and the diffusion coefficients,  
the Peclet number assigned to the motion of the particle is a 
constant and does not vary
in time, i.e.,
\[
 Pe=\frac{v_0(t)}{\sqrt{D_R(t) D_T(t)}}= \frac{v_0}{\sqrt{D^{\star}_R D^{\star}_T}}. 
\]
 With these assumptions, the rotational diffusion 
coefficient and the motility are synchronized and,  consequently,  
the  persistence  length of the ABP remains constant. Thus, on a coarse-grained level,  i.e., at times longer than the characteristic time of the environmental  fluctuation,  the 
fluctuating model exactly reproduces the dynamics of the
ordinary ABP model. 

 An equivalent representation of Eq.~(\ref{mle}) is provided by the Fokker-Plank equation for the PDF of the particle displacement and 
 orientation, $P(\bm{r},\phi,t)$, which is given by
  \begin{eqnarray}
 \partial_t P(\bm{r},\phi,t)&=& - \varUpsilon(t)v^{\star} \hat{\bm{e}}(t) \cdot {\nabla} P+ \varUpsilon(t)D^{\star}_T \Delta P\nonumber \\
&+&\varUpsilon(t)D^{\star}_R\partial^ 2_\phi  P~,
 \label{fokkir}
  \end{eqnarray}
  with $\bm{\nabla}=(\partial_x, \partial_y)$ and $\Delta =
  \frac{\partial^2}{{\partial x}^2}+ \frac{\partial^2}{{\partial y}^2}$.  Equations~(\ref{mle}) and (\ref{fokkir}) provide our toolbox for
  studying the motion of an ABP in an evolving environment whose dynamics   is governed by the stochastic process 
 $\varUpsilon(t)$. 
\section{Theoretical results for a single particle} 
 \label{Theory-results}
  In what follows, we perform a through analysis of the dynamics of the particle based on the Fokker-Plank equation Eq.~(\ref{fokkir}). To this end, we employ the  subordination 
 method~\cite{bochner2005harmonic}.

\subsubsection{Subordination}
\label{subi}

{Subordination is a powerful mathematical technique to treat complex stochastic processes \cite{bochner2005harmonic}.
 The essence of the subordination method is to associate  a random variable  with  the time unit of the subordinated process 
 (see Refs.~\cite{weron2009anomalous,thiel2013disentangling} for examples).} Recently, Chechkin \textit{et al.} \cite{chechkin2017brownian}
extended this concept to the problem  of a passive Brownian particle with diffusing diffusivity. This was done  by connecting the diffusivity  to the random time increment of the original Brownian motion.
Inspired by this work, we rewrite the Eq.~(\ref{fokkir}) in the subordinated  form 
 \begin{subequations}
\begin{align}
\partial_u P(\bm{r},\phi,u)&= - v^{\star} \hat{\bm{e}}(u) \cdot {\nabla} P+ D^{\star}_T \Delta P
+D^{\star}_R\partial^ 2_\phi  P,  \\
\frac{\partial u}{{\partial t}}&= \varUpsilon(t),
\end{align}
 \label{fokkiri}
\end{subequations}
 
where we introduced a new, random variable $u$ whose time derivative is given by $\varUpsilon(t)$.  We refer to $u$ as the subordinator. 

The Fokker-Plank equation ~(\ref{fokkiri}) corresponds to the set of 
Langevin equations  

\begin{subequations}
\begin{align}
\dot{\textbf{r}}(u)&=v^{\star} \hat{\bm{e}}(u) +\sqrt{2 D^{\star}_T} \bm{\xi}_{T}(u), \\
\dot{\phi}(u)&=\sqrt{2 D^{\star}_R} \xi_R(u),  \\
\frac{\partial u}{{\partial t}}&= \varUpsilon(t)~.
\end{align}
 \label{langsub}
\end{subequations}

In the next paragraph, we will use Eqs.~(\ref{fokkiri}) and (\ref{langsub}) to calculate the PDF of particle displacement (and its moments).

\subsubsection{Distribution of displacement and its moments}
\label{momis}
 {As demonstrated in the previous section, the subordination technique allows one to transform the complicated (due to the stochastic parameter $\varUpsilon(t)$ ) form of the Fokker-Plank equation~(\ref{fokkir}) into a set of  simpler equations given in Eqs.~(\ref{fokkiri}). Indeed,  Eq.~(\ref{fokkiri}a) has  the  form of the Fokker-Plank equation for an ordinary ABP where the diffusivities and 
the motility are constant~\cite{sevilla2015smoluchowski}, however, the regular time variable  $t$ is now replaced by  a random variable $u$ in Eq.~(\ref{fokkiri}b).   Therefore, formally the  PDF  $G(\textbf{r},\phi,u)$, satisfying   Eq.~(\ref{fokkiri}a) has the same form as the PDF of an ordinary ABP. The subordinator $u$ is calculated by $u(t)= \int_0^t \varUpsilon(t')dt'$ whose PDF is denoted by $U(u,t)$. To obtain $P(\textbf{r},\phi,t)$, one has to perform an average of  $G(\textbf{r},\phi,u)$ over all the possible  subordinator values with the corresponding probabilities. Therefore one has   
  \begin{equation}
 P(\textbf{r},\phi,t)=\int_0^\infty U(u,t) G(\textbf{r},\phi,u)du~.
  \label{fo1}
\end{equation}
 Here we  focus on the  coarse-grained PDF involving only the particle displacement. Integrating
 both sides of  Eq.~(\ref{fo1}) over $\phi$, one obtains
\begin{equation}
  P(\textbf{r},t)=\int_0^{2\pi} d\phi P(\textbf{r},\phi,t) =\int_0^\infty du U(u,t) G(\textbf{r},u) \nonumber, 
\end{equation}
where $G(\bm{r},u)$ is the coarse-grained PDF of an ordinary ABP as a function of the subordinator $u$. 

 Fourier transformation yields
 
   \begin{eqnarray}
\hat{P}(\bm{k},t) &=& \int_0^\infty U(u,t) \hat{G}(\bm{k},u)du~. 
  \label{fo1k}
\end{eqnarray}

An analytical  form of  $G(\textbf{r},u)$ and  $\hat{G}(\textbf{k},u)$  over the whole span of  $u$ is not 
available, which hinders the calculation of the full function $P(\textbf{r},t)$. However, one can obtain the   asymptotic, long-wavelength,  behavior by inserting $ \hat{G}(\bm{k},u)$ for  corresponding results for the Fourier 
transform of the PDF of displacement of an ordinary ABP. From 
Ref.~\cite{sevilla2015smoluchowski}, it is known that $\hat{G}(\textbf{k},u)$ has the following form:
  $\hat{G}(\bm{k},u)\approx e^{-D'k^2 u}$ as $k \frac{{v^{\star}}}{ D^{\star}_R }\rightarrow 0$ with
  $D'=\frac{{v^{\star}}^2}{2 D^{\star}_R }+D^{\star}_T$ and $k=|\bm{k}|$.  Inserting this result into Eq.~(\ref{fo1k}) we obtain
  
 \begin{eqnarray}
\hat{P}(\bm{k},t) &\approx& \int_0^\infty U(u,t)e^{-D'k^2 u} du\nonumber \\ 
 &=& \hat{U}({D'\bm{k}}^2,t)
 \label{f22f}
\end{eqnarray} 
with $\hat{U}$ being the Laplace transform, that is $\hat{U}(s,t)=\int_0^\infty U(u,t) e^{-su} du$ with $s=D'k^2$.  We recall that Eq.~(\ref{f22f}) has been obtained under the assumption   
$k \frac{{v^{\star}}}{ D^{\star}_R }\rightarrow 0$. This  implies  that the result for $P(\bm{r},t)$ obtained by taking the  inverse Fourier transform
 will be valid only at distances larger than the persistence length, i.e.  for $r \gg \frac{{v^{\star}}}{ D^{\star}_R} $.

 The aforementioned long-wavelength approximation is not required if one focuses only on the moments of  function $P(\bm{r},t)$. To this end, we recall   that the $\textit{mth}$ moment of the PDF can be calculated 
  from the relation  $\langle \bm{r}^m(t)\rangle= (-i)^n {\nabla}^m_{\bm{k}}  \hat{P}(\bm{k},t)|_{\bm{k}=0}$. 
Applying this relation on Eq.~(\ref{fo1k}), and defining 
    $\langle \bm{r}_{OABP}^m(u)\rangle= (-i)^n {\nabla}^m_{\bm{k}}  \hat{G}(\bm{k},u)|_{\bm{k}=0} $    one  arrives  at

  \begin{equation}
   \langle \bm{r}^m(t)\rangle=\int_0^\infty U(u,t) \langle \bm{r}_{OABP}^m(u)\rangle du,
 \label{momtrans}
\end{equation} 

where the subscript $OABP$ indicates that the moments stem from the PDF of an ordinary ABP. Equation~(\ref{momtrans}) is an exact relation between the positional moments of the fluctuating and those of the ordinary ABP model. Here we are interested in the second moment which equals the mean squared displacement~(MSD) by setting the initial position to zero. For an ordinary ABP it is given by \cite{sevilla2015smoluchowski}

   \begin{eqnarray}
  \langle \bm{r}^2_0(u)\rangle&=&\frac{4{v^{\star}}^2}{{D^{\star}_R}^2} \Big[\Big(\frac{D^{\star}_R D^{\star}_T}{{v^{\star}}^2} + \frac{1}{2}\Big)D^{\star}_Ru \nonumber \\
  &-&\frac{1}{2}\Big(1 - e^{-{D^{\star}_Ru}}\Big)\Big], 
 \label{momoab2}
\end{eqnarray} 

where $u$ here plays the role of time. 
 This  MSD exhibits two crossover: the first occurs from diffusive behaviour at short times to a ballistic regime due to  activity-induced directed motion.  The second  crossover  to a another 
 diffusive regime occurs at times longer than rotational relaxation time, where the directed motion is randomized.

 Inserting  Eq.~(\ref{momoab2}) into  Eq.~(\ref{momtrans}), we obtain for the MSD of the fluctuating ABP 
 
 \begin{eqnarray}
\langle \bm{r}^2(t)\rangle &=& \frac{4{v^{\star}}^2}{{D^{\star}_R}^2} \Big[ \Big(\frac{D^{\star}_R D^{\star}_T}{{v^{\star}}^2} + \frac{1}{2}\Big)D^{\star}_R  \Big(\int_0^\infty U(u,t) udu \Big)\nonumber \\
&-& \frac{1}{2}\Big( \int_0^\infty U(u,t)du-  \int_0^\infty U(u,t) e^{-uD^{\star}_R} du \Big) \Big] \nonumber \\
&=& \frac{4{v^{\star}}^2}{{D^{\star}_R}^2} \Big[ \Big(\frac{D^{\star}_R D^{\star}_T}{{v^{\star}}^2} + \frac{1}{2}\Big)D^{\star}_R \langle u(t)\rangle \nonumber \\
&-& \frac{1}{2}\Big( 1- \hat{U}(D^{\star}_R,t) \Big)\Big].  
\label{2momtrans}
 \end{eqnarray}

 The  first integral on the right side of Eq.~(\ref{2momtrans}), $ \langle u\rangle= \int_0^\infty U(u,t) udu$, which represents the first moment of ${U}$, is calculated  using the  first derivative of the Laplace transform $\hat{U}(s,t$) with
respect to $s$ at \mbox{$s=0$}, i.e. $ \langle u\rangle=-\frac{\partial \hat{U}(s,t)}{\partial s} \mid_{s=0} $.
The result of the second integral is unity, due to the normalization of the PDF. Finally  the third integral $\int_0^\infty U (u,t) e^{-uD^{\star}_R} du$ equals  
 essentially   the local 
 value of the Laplace transform of
 the kernel at  $s={D}^\star_R $. In a same manner, one can continue
these calculations for higher moments  to study the impact of the fluctuations on the dynamics.

%
%
%

 \subsubsection{Explicit results}
 \label{sec:fluc}

  So far we did not specify the  process $\varUpsilon(t)$, which encodes all the information regarding the fluctuations of the environment, see Eqs.~(\ref{mle}).
Following the suggestions by Jain et al. \cite{jain2016diffusion} and Chechkin et al. \cite{chechkin2017brownian},  we consider the 
 $\varUpsilon(t)$ to be the square of 
 a $n$-dimensional Ornstein-Uhlenbeck process, i.e. $\varUpsilon(t)=\textbf{Y}_n^2(t)$, where $\textbf{Y}_n$ can be interpreted as the position 
 vector of a $n$-dimensional harmonic oscillator that performs
Brownian motion. With this in mind, the process $\varUpsilon(t)$ is determined by the  Langevin-like equations,
\begin{eqnarray}
\varUpsilon(t) &=& {\textbf{Y}_n^2(t)} \nonumber \\ 
\dot{\textbf{Y}}(t)&=& -\frac{1}{\tau_o}\textbf{Y}(t)+\sigma \bm{\zeta}(t),
\label{ou}
\end{eqnarray}
where $\bm{\zeta}$ is a Gaussian white noise with zero mean and  correlation function  
$\langle \zeta_i(t_1)\zeta_j(t_2) \rangle=\delta_{ij}\delta(t_1-t_2)$ for $i,j=1,...,n$. The parameters  $\sigma$ and $\tau_o$ denote the relaxation time and noise strength of three dimensional (auxiliary) variables $\bm{Y}$ with physical dimensions 
 $[\sigma]={s^{-1/2}}$ and $[\tau_o]=s$.   Fluctuations become dominant for $t \ll \tau_o$.

\textit{Equilibrium initial conditions}: In the following, we present results for the dynamics starting from two different initial conditions for the process $\bm{Y}$. These are particularly important for short times~\cite{sposini2018random}.
Equilibrium initial conditions describes a situation where the measurement starts long after the process  $\textbf{Y}$ has reached its
stationary state. To model this situation, the initial value   $\textbf{Y}_0$ is
randomly  taken from the equilibrium Boltzmann distribution of the  process $\textbf{Y}$. 
The mean value of  $\varUpsilon$ is then determined by the stationary   auto-correlation function of $\bm{Y}_{st}$, 
  $\langle \varUpsilon_{st} \rangle=\langle\textbf{Y}^2_{st}\rangle=\frac{n\sigma^2 \tau_o}{2}$ for $t\gg\tau_o$.
 For such a choice of the initial conditions, 
the equilibrium PDF, ${U}^{eq}_n(u,t)$, of the process  $u(t)= \int_0^t \varUpsilon(t')dt'$  is given  in the Laplace domain~\cite{dankel1991distribution} by

  
\begin{eqnarray}
 \hat{U}^{eq}_n(s,t)=\exp{(\frac{nt}{2\tau_o})}/\Big[\frac{1}{2}(\sqrt{1+2s\sigma^2 \tau^2_o}+\frac{1}{\sqrt{1+2s\sigma^2 \tau^2_o}}) \nonumber \\
 \times\sinh(\frac{t}{\tau_o} \sqrt{1+2s\sigma^2 \tau^2_o}) +\cosh(\frac{t}{\tau_o} \sqrt{1+2s\sigma^2 \tau^2_o}) \Big]^{n/2}~. \nonumber \\
 \label{cg2eq}
\end{eqnarray}
The quantity  $ \langle u\rangle$ can now be readily calculated as \mbox{$\langle u\rangle=-\frac{\partial \hat{U}^{eq}_n(s,t)}{\partial s} 
\mid_{s=0} = \frac{n\sigma^2 \tau_o}{2}t$} which leads to the following expression for the 
MSD of the fluctuating ABP

 \begin{eqnarray}
\langle \bm{r}^2(t)\rangle &=& 4 \bar{D}_Tt + 2 \frac{\bar{v}^2}{{\bar{D}^2_R}}\Big(\bar{D}_Rt-1+\hat{U}(D^{\star}_R,t)\Big)~. \nonumber \\
\label{2momtrans2}
 \end{eqnarray} 
In Eq.~(\ref{2momtrans2})  $\bar{D}_T=\frac{n\sigma^2 \tau_o}{2} D^{\star}_T$, $\bar{D}_R=\frac{n\sigma^2 \tau_o}{2}D^{\star}_R$ and $\bar{v}=\frac{n\sigma^2 \tau_o}{2}v^{\star}$ represent 
the stationary values of the diffusivities and the motility.

 Fig.~\ref{msds-single} shows the  MSD   according to Eq.~(\ref{2momtrans2}) for the set of
 parameters of  $v^\star=D^\star_R=\sigma=\tau_o=1$ and $D_T^\star=0.1$, and for two values of $n$, together with 
 the corresponding simulation results. To this end, we here solved  Eqs.~(\ref{sync}) and (\ref{mle}) with the Euler's first-order method. The simulation details can be found in Sec.~\ref{sec:resultat}. As seen in
 Fig.~\ref{msds-single}, the  
behavior  of the MSDs at very short and long times (with respect to $\bar{\tau}_R=\frac{1}{\bar{D}_R}$)  matches with corresponding long- and short time 
behavior of the MSD of the conventional ABP, that
is, linear growth of the MSD with respect to time. This finding, that the linear part of MSD  remains unchanged under fluctuations, is in  
agreement with the case of the passive
Brownian particle, widely studied in the context of diffusing diffusivity problem. At   intermediate times, however,   a slight difference
between the 
MSD of the fluctuating and that of the conventional ABP is observed. Qualitatively, one observes that the ballistic behavior of the MSD of the conventional ABP is 
replaced by a smooth cross-over when the fluctuations are introduced. The difference  between two MSDs is, however, rather small and can be neglected
in the presence of whatever experimental 
noise. The  ratio of these two MSDs is plotted as inset in Fig.\ref{msds-single}. This graph suggests that, for higher dimensions of the 
subordinator process, $n>1$, the observed difference  becomes
even less significant. 

\begin{figure}[h!]\center
{\includegraphics[width=0.48\textwidth]{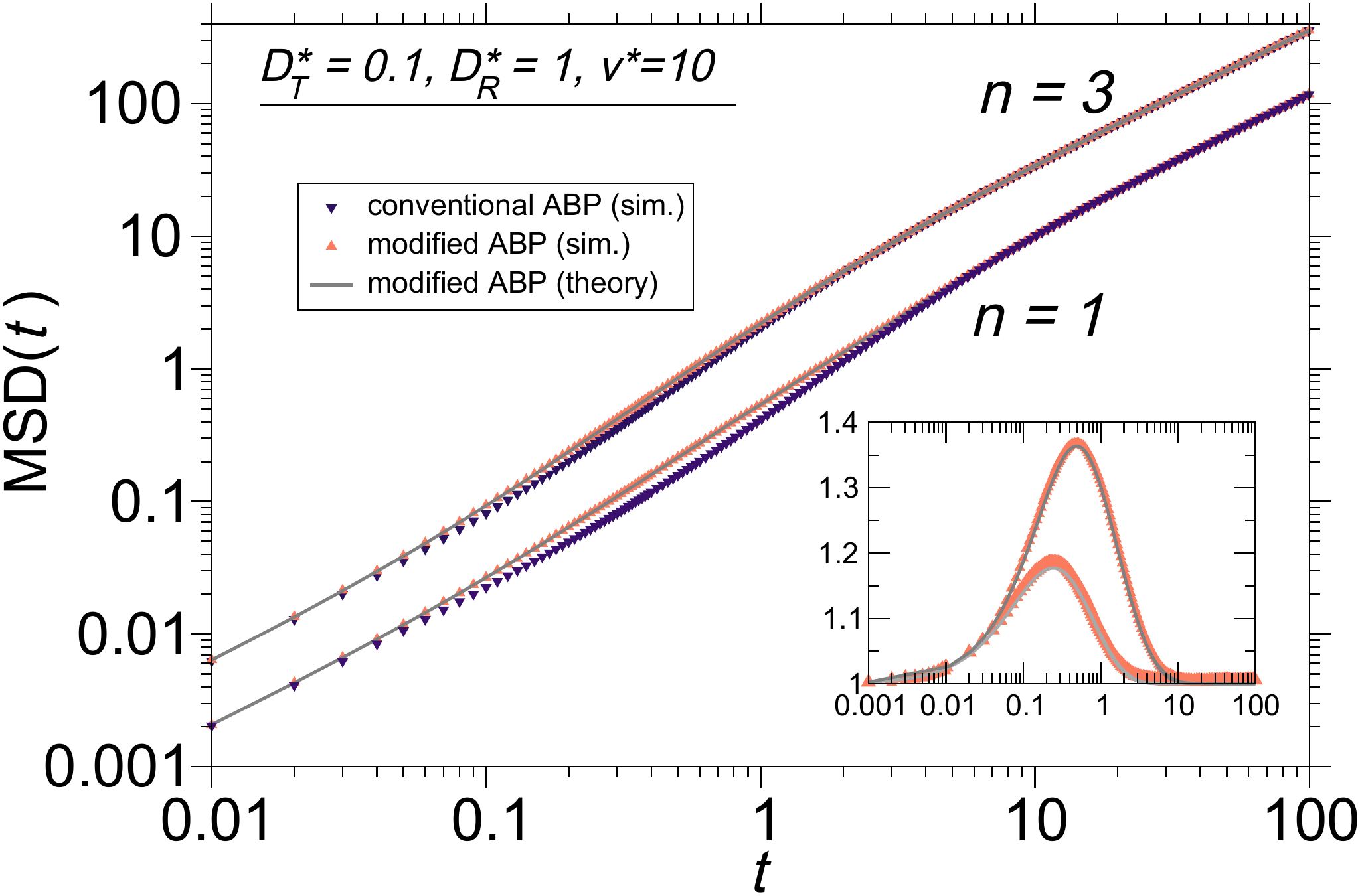}}
\caption{Comparison of the MSDs for the ordinary ABP and the modified one with different  dimension $n$ for Ornstein-Uhlenbeck process. 
The parameters used here are $v^\star=1$, $D_R^\star=1$, and $D_T^\star=0.1$. The {qualitative behavior} of the  MSDs does not change in presence of  fluctuations. 
The short and long time behavior of MSDs are in quantitative agreement, although a small quantitative deviation at intermediate times (the ballistic regime) is observed. 
The inset shows the ratio of the MSDs of the modified ABPs and the conventional one. }
\label{msds-single}
\end{figure}

 As we proceed to show, the  MSD changes when the time scales corresponding to the fluctuations and
 rotational relation time deviate from each other.  In Fig.~\ref{fig:msd-eic}, we plot the MSD of 
 the fluctuating  ABP according to Eq.~(\ref{2momtrans2}) for different values of $\bar{D}_R=\frac{1}{\bar{\tau}_R}$ (through changing the value of $D^{\star}_R$ ) to fulfil the following limiting cases:  i) $\bar{\tau}_R \ll \tau_o$, ii) $\bar{\tau}_R =\tau_o$ and iii) $\bar{\tau}_R\gg \tau_o$. The first 
 two cases are very similar and contain no ballistic regime. However,  the emergence of a ballistic regime 
 at times $\tau_o \ll t \ll \bar{\tau}_R $ is evident in the latter case. 
 \begin{figure}[ht]
   \centering
   \includegraphics[width=0.45\textwidth]{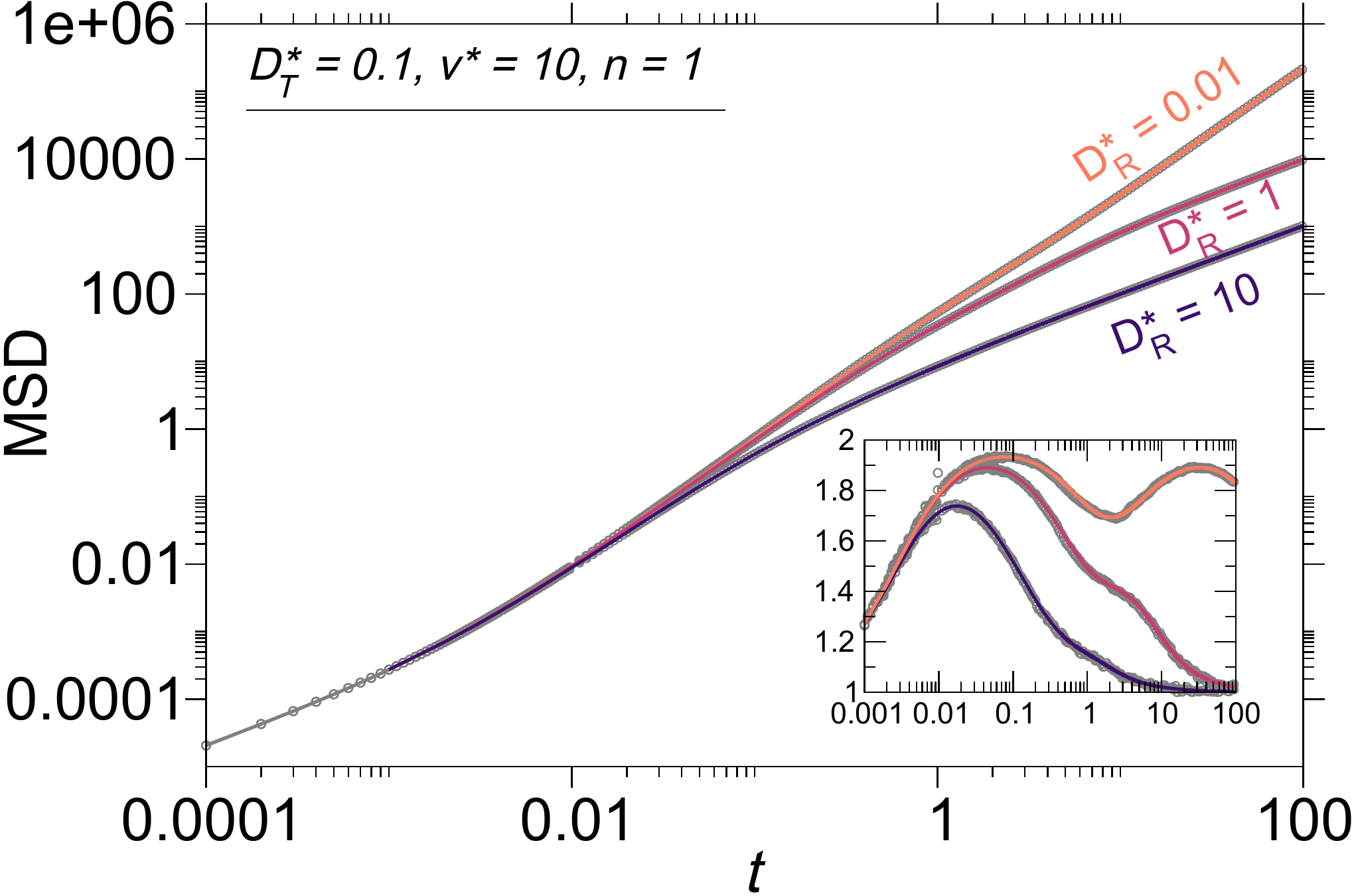}
   \caption{\label{fig:msd-eic}Comparison of the MSD of the fluctuating ABP with different $D^\star_R$ values (with the equilibrium initial conditions). The gray symbols represent the simulation results and the lines represent the predictions of \equref{2momtrans2}. The inset shows the exponent of the MSD calculated as $\alpha=d \log{\mathrm{MSD}}/d \log{t}$ as a function of time. }
 \end{figure}

 In what follows, we investigate the asymptotic behavior  of the MSD for the  case when ${\bar{D}}_R\gg 1/\tau_o$. To do so, we  separate the relevant time 
 scales in the   ${U}^{eq}_n(D^{\star}_R,t)$.  For ${\bar{D}}_R\gg 1/\tau_o$, using the Taylor expansion
 as $\frac{t}{\tau_o}\sqrt{1+\frac{4\bar{D}_R\tau_o}{n}}\simeq \frac{t}{\tau_o}+\frac{2\bar{D}_Rt}{n}$, for the arguments of $\sinh$ and $\cosh$ functions in \equref{cg2eq} and with some algebra 
 we can
obtain the asymptotic behavior of the MSD in different time-scales for $n=1$ as follows:

\begin{equation}
 \langle \bm{r}^2(t)\rangle\sim\begin{cases}
   4 \bar{D}_Tt & \text{ $t \ll \tau_o \ll \bar{\tau}_R$ }\\
   \bar{v}^2 t^2  & \text{$ \tau_o\ll t \ll \bar{\tau}_R$}\\
  4( \bar{D}_T +  \frac{\bar{v}^2}{{2\bar{D}_R}})t   & \text{$ \tau_o\ll\bar{\tau}_R  \ll t$}.
  \end{cases}
  \label{msd-d-eq}
\end{equation}
 According to Eq.~(\ref{msd-d-eq})~ the well-known behavior of the MSD of the ordinary ABP  (see Eq.~(\ref{momoab2})) is preserved when 
 the fluctuations are introduced in the equations of the motion. Nevertheless,
 the fluctuations only affect the cross-over from the linear regime at short times to the ballistic regime in the intermediate times. Our result here is complementary to the case, where the fluctuations arise internally which results in 
 more complex dynamics~\cite{23inlutz}.

%
%

 We now proceed  by investigating the behavior of the PDF of particle displacement. Considering the PDF of the sub-ordinator 
 process as in Eq.~(\ref{cg2eq}), and  using Eq.~(\ref{f22f}) we obtain the asymptotic behavior of the PDF for $t\gg \bar{\tau}_R$  as 
 \begin{eqnarray}
\hat{P}(\bm{k},t) \sim \hat{U}^{eq}_n({D'\bm{k}}^2,t).
 \label{f22f-eq}
\end{eqnarray}

Following the  discussion in Sec.~\ref{momis}, our  asymptotic result for the PDF of the particle displacement is identical  to the case of a passive Brownian particle under fluctuations~\cite{chechkin2017brownian}, while the diffusion coefficient is substituted with the coarse-grained one being $D'=\frac{{v^{\star}}^2}{2 D^{\star}_R }+D^{\star}_T$. Therefore, our analysis of the PDF is relevant only at times $t>{\tau}^\star_R$.
  Nevertheless, if the characteristic time for 
 the environmental fluctuations is longer than ${\tau}^\star_R$, one can still capture the effect of fluctuations on the behavior of the PDF. In  Fig.~\ref{pdfs-single}, we compare
 the form of the PDF for the conventional and the fluctuating ABPs. The form of the PDF for  the conventional ABP
 remains Gaussian at all times. However, an exponential behavior of the PDF of displacement for the 
 fluctuating ABP is observed at short times, despite the linear behavior of the MSD. This is the signature of the diffusing diffusivity model. For 
 times longer than $\tau_o$ the emergence of the 
 Gaussian form of the PDF is evident. For a comprehensive asymptotic analysis of the PDF of particle displacement in the passive case, we refer the reader to Ref.~\cite{chechkin2017brownian}.

  \begin{figure}[h!!]\center
    {\includegraphics[width=0.45\textwidth]{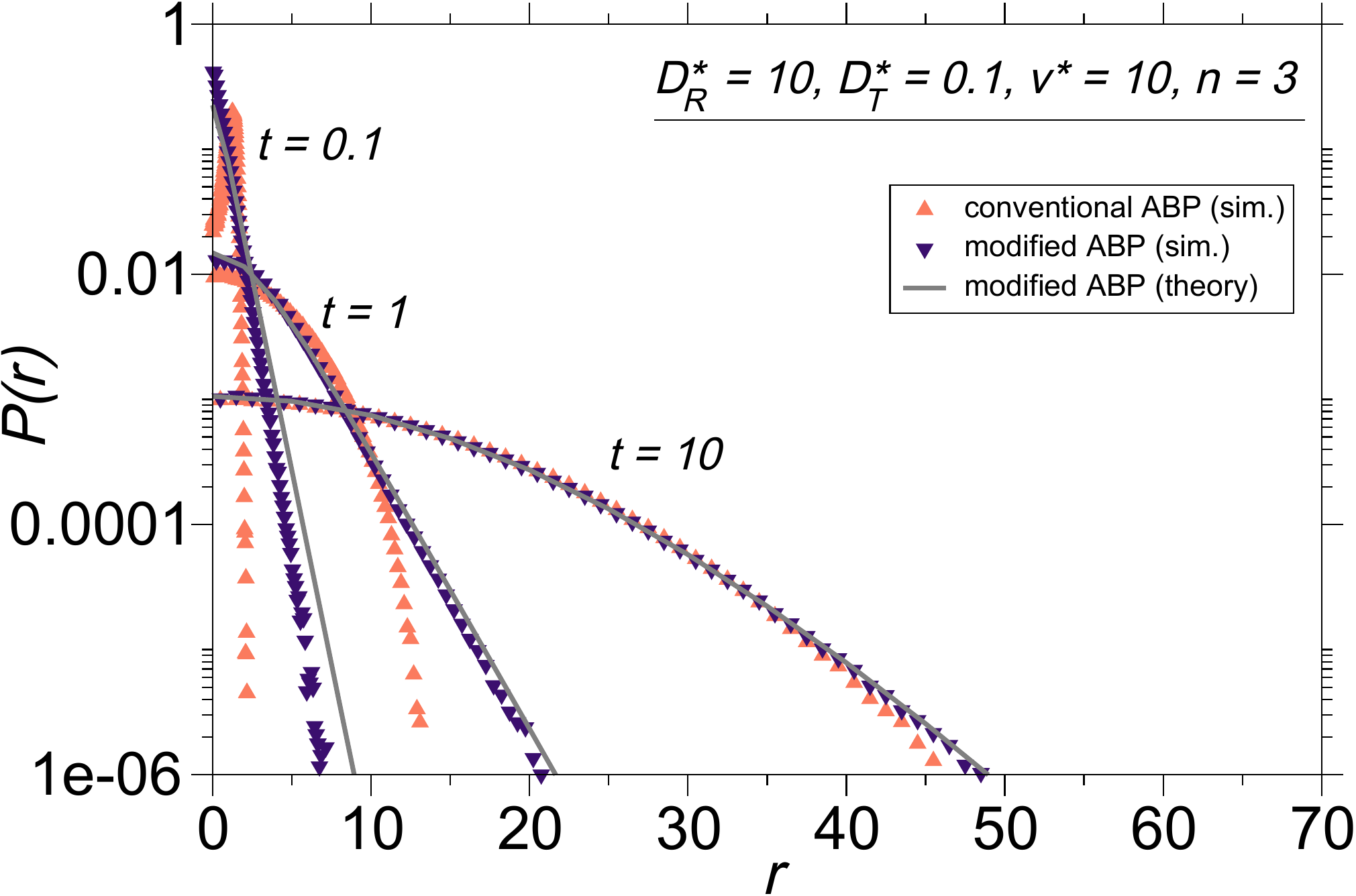}}
\caption{Comparison of PDFs of the conventional ABP model  and the modified one, {corresponding to the MSDs depicted in \figref{msds-single}} with $n=3$.  }
.\label{pdfs-single}
\end{figure}

\textit{Non-equilibrium initial conditions}: We now consider a situation where the process $\textbf{Y}$ at $t=0$ is not in its 
stationary state. For an example, one could imagine that the measurement starts just when the particle is inserted into the medium.  Then, the PDF of the particle displacement  differs at short times from the corresponding PDF for
the equilibrium initial condition.  Nevertheless, as  time evolves, the  process $\textbf{Y}$ reaches its stationary state and one expects 
an identical behavior of the PDFs. Here, without loss of the generality, we assume  the
initial condition  $\textbf{Y}_0=0$. In this case,  the  non-equilibrium PDF, ${U}^{neq}_n(u,t)$, for the process  $u(t)= \int_0^t \varUpsilon(t')dt'$ in the Laplace
domain  reads~\cite{dankel1991distribution}
 \begin{eqnarray}
&\hat{U}&^{neq}_n(s,t)=\exp{(\frac{nt}{2\tau_o})}/\Big[\frac{1}{\sqrt{1+2s\sigma^2 \tau^2_o}} \nonumber \\
&\times&\sinh(\frac{t}{\tau_o} \sqrt{1+2s\sigma^2 \tau^2_o})+\cosh(\frac{t}{\tau_o} \sqrt{1+2s\sigma^2 \tau^2_o}) \big]^{n/2}, \nonumber \\
 \label{cg2neq}
\end{eqnarray}
whose first moment is  $ \langle u\rangle=\frac{n\sigma^2 \tau_o}{2}(t-\frac{1}{2}\tau_o(1-e^{\frac{-2t}{\tau_o}}))$. This leads to a complex form for the
MSD which strongly differs from that for an ordinary ABP:

 \begin{eqnarray}
\langle \bm{r}^2(t)\rangle &=& \Big(4 \bar{D}_T + 2 \frac{\bar{v}^2}{{\bar{D}_R}}\Big)\Big(t-\frac{1}{2}\tau_o(1-e^{\frac{-2t}{\tau_o}}\Big) \nonumber \\
&-&2 \frac{\bar{v}^2}{{\bar{D}_R}^2}\Big(1-\hat{U}(D^{\star}_R,t)\Big).
\label{2momtrans2neq}
\end{eqnarray}
Fig.~\ref{msd-d-neic} shows the MSD according to Eq.~(\ref{2momtrans2neq}) for different values of $D^{\star}_R$. 
\begin{figure}[ht]
  \centering
  \includegraphics[width=0.45\textwidth]{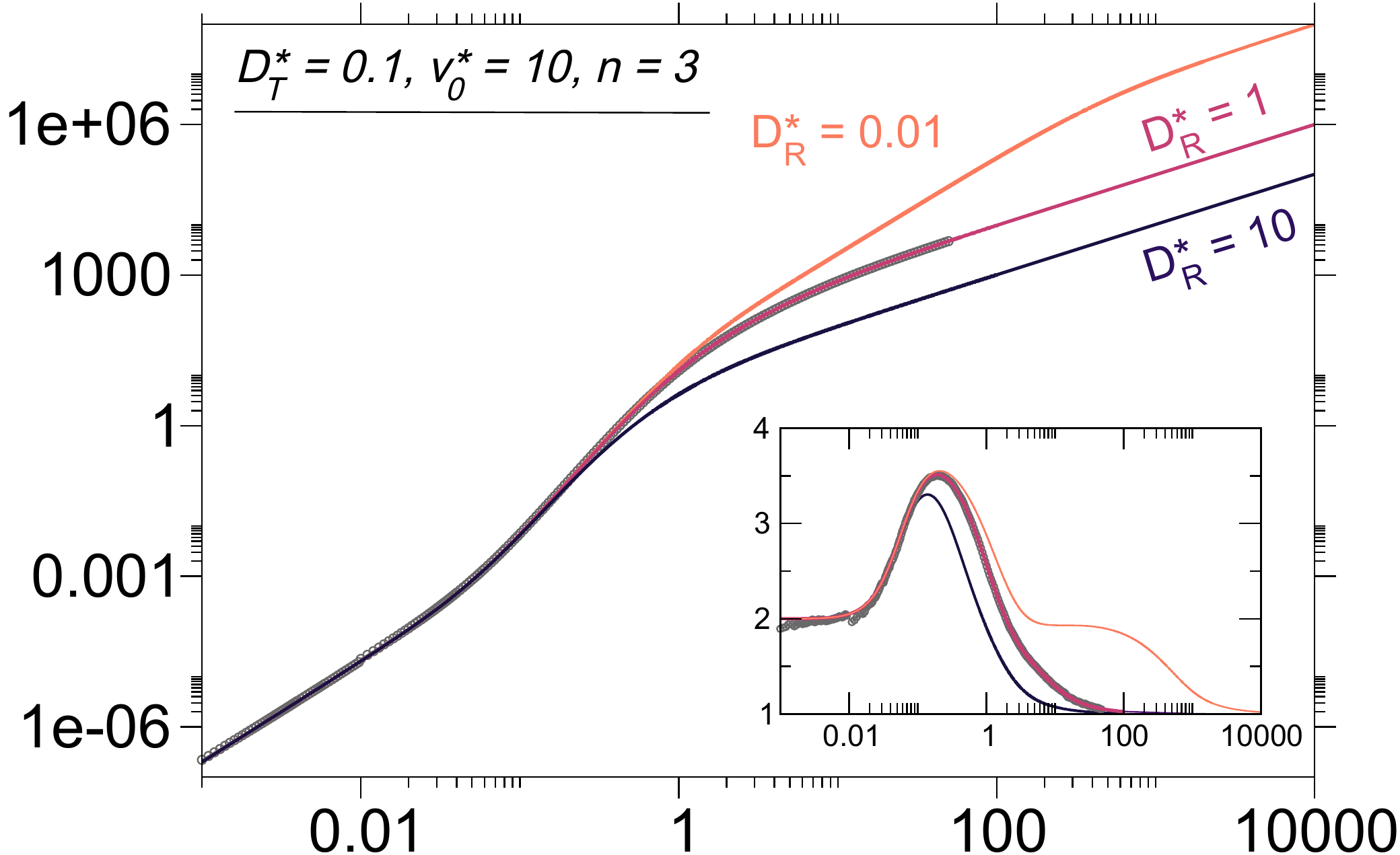}
  \caption{Comparison of the MSD of the fluctuating ABP with different $D^\star_R$ values (with  non-equilibrium initial conditions). The gray symbols represent the simulation results and the lines represent the predictions of \equref{2momtrans2neq}. The inset shows the exponent of the MSD.}
  \label{msd-d-neic}
\end{figure}

In analogy to the case of equilibrium initial condition, 
we perform the asymptotic analysis of the MSD, yielding
\begin{equation}
 \langle \bm{r}^2(t)\rangle\sim\begin{cases}
   4 \bar{D}_T \frac{t^2}{\tau_o}& \text{ $t \ll \tau_o \ll \bar{\tau}_R$ }\\
   \bar{v}^2 t^2  & \text{$ \tau_o\ll t \ll \bar{\tau}_R$}\\
  4( \bar{D}_T +  \frac{\bar{v}^2}{{2\bar{D}_R}})t   & \text{$ \tau_o\ll\bar{\tau}_R  \ll t$}.
  \end{cases}
\end{equation}

Clearly, the dynamics of the MSD at times shorter than $\tau_o$ is now ballistic. This is explained by the initial acceleration due 
to the non-equilibrium initial condition. Apart from that, 
at longer times, $t\gg\tau_o$, the expected ballistic behavior arising from the persistent motion of the ABP reappears, as the times scales associated with the
environmental fluctuations
and the rotational relaxation deviate from each other. At even longer times $ t \gg \tau_o \gg \bar{\tau}_0$  linear behavior of the MSD at times   emerges.

We further proceed by calculating the PDF of the particle displacement according to Eq.~(\ref{cg2neq}), 
\mbox{$\hat{P}(\bm{k},t) \sim \hat{U}^{neq}_n({D'\bm{k}}^2,t)$},  at times longer and equal  to the characteristic time for the environmental fluctuations $(t_o=1)$. As seen in Fig.~\ref{pdfs-neic},  the PDF at long times $(t\gg t_o)$ has a  Gaussian form which exhibits  a strong non-Gaussianity at shorter times $(t\sim t_o)$ which has widely been discussed in 
\citep{sposini2018random}. We note that the agreement between the theory and simulation results is expected, as we have 
chosen the small values for the rotational relaxation time. 
\begin{figure}[ht]
  \centering
  \includegraphics[width=0.45\textwidth]{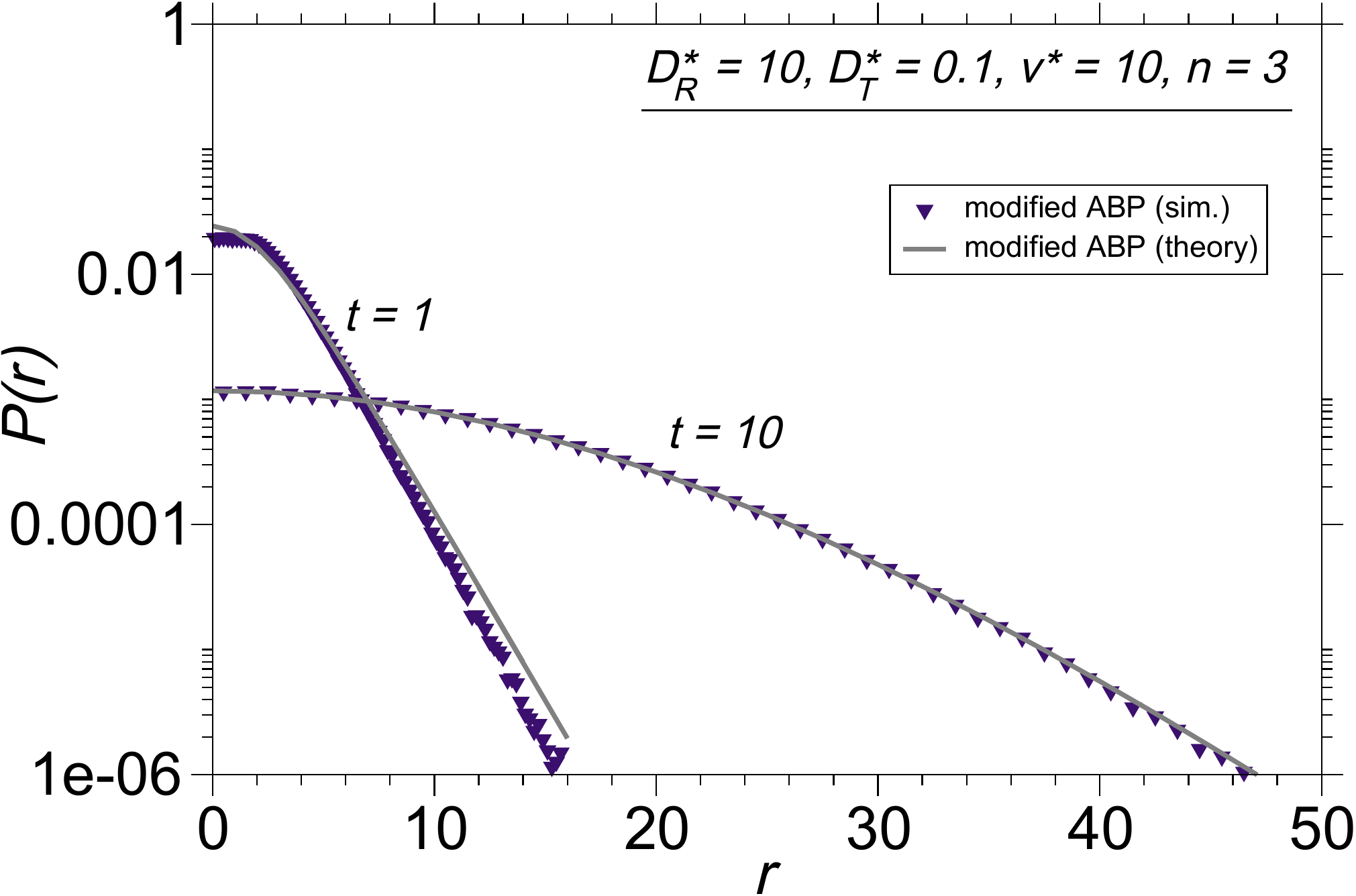}
  \caption{PDF of the particle displacement at times longer and equal to the characteristic time for the fluctuations.}
    \label{pdfs-neic}
\end{figure}
 
   We finally note that the behaviour of the PDF of particle orientation, $P(\phi,t)$, is independent of the activity. Under the aforementioned fluctuations with an equilibrium initial conditions, it was shown~\cite{jain2017diffusing}  that $P(\phi,t)$ exhibits 
  exponential behaviour at times shorter than the  characteristic time of the fluctuations and a Gaussian behaviour at longer times is recovered.   In what follows, we will study the collective behaviour of an ensemble of  fluctuating ABPs. Then, we  will show how these short times behaviours  of PDFs of the rotational and translational displacements  can  alter the dynamic of the system.


\section{Collective Behavior}\label{sec:resultat}
It is wellknown that  active particles show intriguing collective phenomena   including self-ordered motion~\cite{vicsek1995novel}, motility-induced phase separation~\cite{stenhammar2013continuum,buttinoni2013dynamical}, dynamic clustering in presence~\cite{theurkauff2012dynamic} and absence of chemotaxis~\cite{stenhammar2013continuum,buttinoni2013dynamical}, and mesoscale turbulence~\cite{wensink2012meso,reinken2018derivation,reinken2019anisotropic}, to name just a few. Here we focus on the well-studied phenomenon of motility-induced phase separation~(MIPS). There are a number of earlier studies where fluctuations of some sort have been investigated~\cite{weber2016binary,fischer2019aggregation,van2019interparticle}.  Among these studies, Ref.~\cite{fischer2019aggregation} is  most similar to ours, as in that study a non-constant motility was considered. Nevertheless, in Ref.~\cite{fischer2019aggregation} the motility depends on the inter-particle forces such that there is an enhanced  positive feedback between an increase of the  density and the reduction of motility. Contrary to that, here we consider variations of the motility as a direct consequence of fluctuations of the environment in which the particles are embedded.}

\subsection{Simulation setup}

As a special case of a many-particle system of ABP with fluctuations, we first investigate  the infinite dilution limit, i.e. the case where the interactions between the particles are negligible. In this limit, which we refer to as the (effective) single particle case, the dynamics of the particles with positions $\bm{r}_i(t)$, $i=1,...,N$ are governed by \equrefs{mle}. With $\varUpsilon(t)$ given by Eqs.~(\ref{ou}), these equations are numerically  integrated via a  first-order forward Euler scheme with
time-step of $\delta t=10^{-3}$ for $N\sim 10^{6}$ non-interacting particles on a  two-dimensional plane. Particles are initially placed at the center of the plane. Then, the  following equilibrium and non-equilibrium initial conditions for $\textbf{Y}$ are assigned to the particles:  For the equilibrium case,  an Ornstein-Uhlenbeck process  is simulated for each particle for a time range of $10\tau_0$. The obtained $\textbf{Y}$ values are then used as the initial conditions. For the non-equilibrium initial conditions (see Sec.~\ref{sec:fluc}), the initial conditions $\textbf{Y}=0$ is used for all particles. Starting from these initial conditions, \equrefs{mle} are integrated to obtain $\langle \bm{r}^2(t)\rangle$
 and $P(\bm{r},t)$. The results are shown in Figs.~(\ref{msds-single}-\ref{pdfs-neic}) together with corresponding analytical calculations.  

 To study the collective behavior of the system, an inter-particle  interaction is introduced in \equrefs{mle}.  A common choice  to model  ABPs interactions is the soft, purely repulsive and isotropic  Weeks-Chandler-Andersen~\cite{weeks1971role} (WCA)
 potential. This is given by
\begin{equation}
 V(\vr) = 4\epsilon\Big((\frac{\sigma}{|\vr|})^6-(\frac{\sigma}{|\vr|})^{12}\Big),
\label{potential}
\end{equation}
 for $|\vr|\leq r_\mathrm{c}=\sqrt[6]{2}$ and zero elsewhere, where $\vr$ is the distance between two particles, $\sigma$ is the diameter of  particle, and $\epsilon$ is the energy scale of the interaction. The equations of motion for the $i$-th particle then  read as
 \begin{eqnarray}
\dot{\textbf{r}}_i(t)&=&v^{\star} \varUpsilon_i(t) \mathbf{\hat{e}}_i(t) +\sqrt{2D^{\star}_T\varUpsilon_i(t)} \bm{\xi}_{i,T}(t) +
{D^{\star}_T\varUpsilon_i(t)\over k_\mathrm{B}T }\mathbf{F}_i, \nonumber\\
\dot{\phi}_i(t)&=&\sqrt{2D^{\star}_R\varUpsilon_i(t)} \xi_{i,R}(t),\label{mle-dense}
\end{eqnarray}
where the vector $\mathbf{\hat{e}}_i(t):=(\cos\phi, \sin\phi)$ is the heading direction of the $i$-th particle, and $\textbf{F}_i$ is the
force on the $i$-th particle due to its WCA interaction with all other particles.

  For the interacting case, we consider a system of $N=1600$ particles in a square box with lengths  $L_x=L_y=60\sigma$, where  periodic boundary 
conditions are implemented along both $x$ and $y$-directions. Following Refs.~\cite{siebert2017phase,buttinoni2013dynamical,fischer2019aggregation,liao2018clustering}, we choose $\sigma=1$, and $\epsilon\gg k_\mathrm{B}T$~ such that the repulsion comes close to a  hard-core interaction. Note that the force, $\mathbf{F}_i$, enters the dynamics of the particle with a prefactor $D^\star_T\varUpsilon_i(t)/(k_BT)$ [see \equrefs{mle-dense}].  This prefactor does not play a role for a true hard-core interaction, where the force is either zero (no overlap) or infinite (overlap).  However,  in our case,  this prefactor does play an important role: the particles become effectively softer when $D^\star_T\varUpsilon_i(t)$ becomes small, which is a likely case for  the chosen random process of $\varUpsilon_i(t)$.

To avoid this undesired effect, we set  the repulsive potential $\epsilon$  such that $D^\star_T\varUpsilon_i(t)\epsilon/(k_BT)=100$ for all the particles and for all times. In this way, we avoid the undesired softening of the particles.  It is worth mentioning that with these choice of parameters, our reduced interaction energy $\epsilon/(k_BT)$ is two orders of magnitude larger than in  previous studies~\cite{blaschke2016phase,liao2018clustering,siebert2017phase}. The time integration of the resulting equations is performed using a first-order forward Euler scheme with a much smaller time-step compared to the single particle simulations, $\delta t=10^{-5}$, 
to avoid  overlap situation of the disks~\cite{liao2018clustering}.
 Initially, the particles are positioned on a regular square lattice and the system is simulated for $t=50$ in Lennard-Jones time units to reach its stead state.  All the measurements begin after this transient regime. To obtain reasonable statistics, $N_s=5$ independent samples are considered.

\nhs{To complete this section, we introduce the quantities needed for our main objective, i.e. the study of MIPS.} Most importantly we  analyse  the probability distribution of the local density across the system. \nhs{In the coexisting regime (i.e. in presence of MIPS), the probability distribution of the local density shows a double-peak structure ~\cite{reinken2018derivation,reinken2019anisotropic}, where the coexistence densities correspond to the position of the two maxima}.

   To obtain the local density, we first calculate the area associated with each particle, $A_i$, using
Voronoi tessellation~\cite{rycroft2009voro++}. The local density associated with the $i$-th particle is  $\rho_i=1/A_i$, which is related to the particle-resolved local packing fraction  defined by $\phi_i:=\pi\sigma^2 \rho_i/4$. The probability of having 
local density $\rho$ at any chosen point in the plane, $\vr$, is proportional to the number of particles with $\rho_i=\rho$ weighted by the area associated with each of those particles (i.e. $A_i$). In this way one can obtain a probability distribution of the local density associated with position (and not particles). 
This method is equivalent to the procedure suggested in Ref.~\cite{liao2018clustering} with an infinitesimally small lattice spacing.

   We also use $\rho_i$ to obtain the density dependent of the particle velocity considered in sec.~\ref{dddv}. \modified{The velocities are calculated by the displacement of particles, \mbox{$\mathbf{v}_i(t):=(\mathbf{r}_i(t+\Delta t)-\mathbf{r}_i(t-\Delta t))/(2\Delta t)$}, where $\Delta t$ is chosen to be $0.01$ time units. This interval is sufficiently large to average out  instantaneous thermal fluctuations (i.e. during 20 time steps), but still small enough such that the particles do not move significantly compared to their  size.}

\subsection{Motility-induced phase separation}
\begin{figure}[h]
 \includegraphics[width=0.45\textwidth]{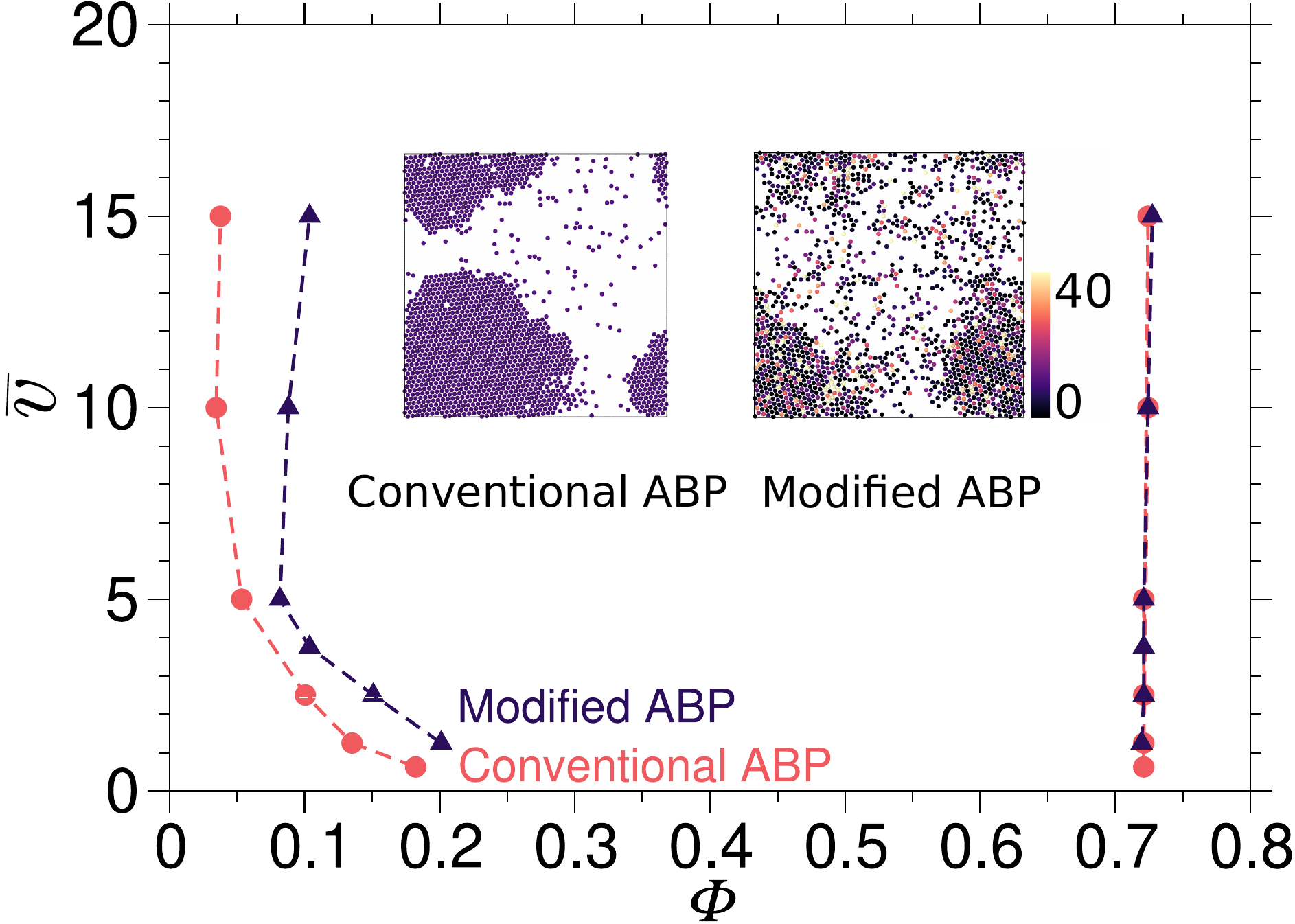}
     \caption{
       {\nhs{Coexistence  densities in ($\bar{v}, \Phi$) plane for} \mbox{$D^\star_R=0.1$} , $D^\star_T=0.001$ , $v^\star=10$, $n=1$, $\sigma_o=1$, at mean area fraction $\bar{\Phi}=0.44$  for the conventional ABP model (red circles) and the fluctuating model (blue triangles). Included  are two simulation snapshots within the phase-separated regime.
The color code in the snapshots reflects   the local density obtained by Voronoi tessellation; the darker color indicates larger density.}
     }
     \label{fig:MPIS-Y1-no-fluctuation}
 \end{figure}

In this section, similar to the Refs. \citep{blaschke2016phase,liao2018clustering}, we investigate  MIPS by obtaining the 
 \nhs{coexistence  local densities in ($\bar{v},\Phi$) plane.} Our main finding is presented in Fig.~\ref{fig:MPIS-Y1-no-fluctuation}  where we compare the coexistence densities for the fluctuating ABP model with those for the conventional one. For illustration, we have included two simulation  snapshots.  The results indicate that introducing fluctuations leads to an increase of the density values on the dilute side, while  the coexistence densities on the high-density side remain unaffected.

Consequently,  the coexistence region is narrower than in the conventional model suggesting that   fluctuations tend to hinder MIPS. This hindrance can understood  as follows. The basic mechanism for  MIPS to occur~\cite{buttinoni2013dynamical}  is that  two particles 'block' each other's path over a time interval which is large enough for the other particles to reach this configuration. This induces a clustering  process~\cite{buttinoni2013dynamical}. The aforementioned time interval  during which   two particles are facing each other can be  obtained from the short time behaviour of the  PDF of particle orientational displacements. While
for  conventional ABPs, this distribution is governed by a Gaussian PDF, for  fluctuating ABP, it was shown in Ref.~\cite{jain2017diffusing} that the short time distribution of the rotational displacements is an exponential function. This means that, in the same time interval, fluctuating ABPs explore larger angles. As a consequence,   collisions  between fluctuating ABPs  are less effective in slowing down the particles, leading to a hindrance of MIPS. As discussed in the next section, this is also in line with our observation that for any local density,  the ABPs have larger velocities in the  fluctuating model as compared to the conventional one.

 So far,  we have discussed the possible microscopic effect of the fluctuations on the cluster formation. Interestingly, the snapshots in Fig.~\ref{fig:MPIS-Y1-no-fluctuation} further  suggest that when the dense phase is formed in presence of 
 fluctuations, the structure of its boundary is less sharp compared to the boundaries of the clusters formed in the corresponding conventional ABP system. This can be explained again by considering the short time behaviour of the 
translational and rotational displacements PDFs.  We focus on the behavior of the particles which are right at the boundary, or, alternatively,  the particles inside the cluster which are close to the boundary. These are the particles which  might find their way out of the cluster.  Roughly speaking, by rotating faster and performing larger displacements (as reflected in the exponential distributions of the rotational and translational displacements at short times), the particles in the fluctuating model find their way out of the cluster faster compared to the ABPs in the conventional model. This explains why the boundary of the dense phase is less sharp.

 \subsection{Density dependent velocities and effective free energy}
\label{dddv}
\begin{figure}[ht]
   \centering
  \includegraphics[width=0.45\textwidth]{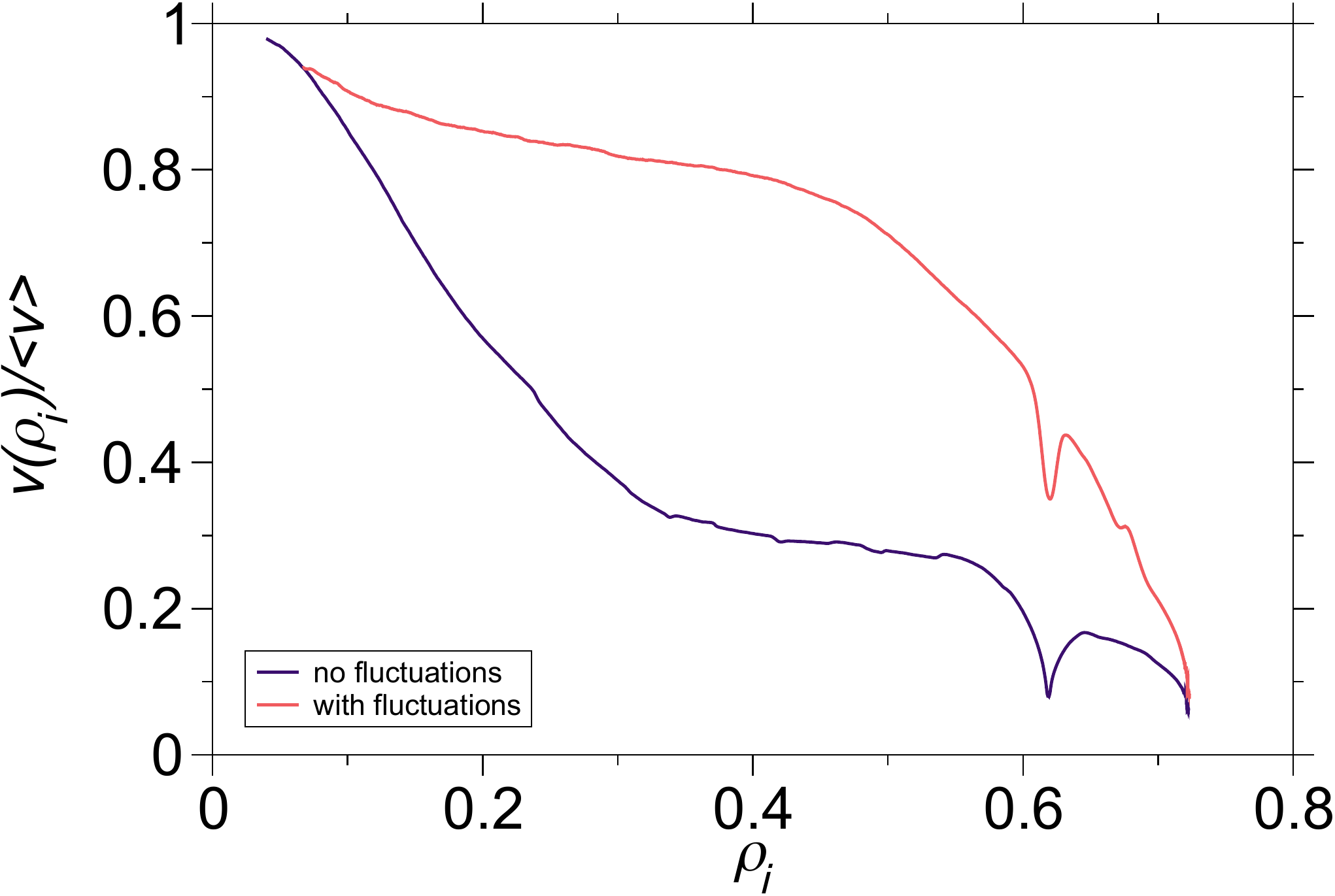}
   \caption{\label{fig:vel-rho}
   Velocity as a function of local particle density for $D^\star_R=0.1$, $D^\star_T=0.001$, $v^\star=10$, \modified{$n=1$} and \nhs{$\bar{\Phi}=0.44$}.}
 \end{figure}

 Further  insight into the effect of the fluctuations on  MIPS can be gained by analyzing the dependence of particle velocities on the local density. This dependency is shown in Fig.~\ref{fig:vel-rho} for systems in presence and absence of fluctuations.

For both systems, $v$ generally decreases with $\rho$ (apart from the local minima at high densities) due to the  increasing packing effect. To better understand the shape of $v(\rho)$, one has to note that the curve shows the velocity averaged over the whole space, 
including  groups of particles with the same local density. In presence of phase separation, where the system displays dilute and dense regimes, the groups of particles with the same $\rho_i$ can have different local environments. In particular, the local minima can be traced back to point defects in a cluster with nearly vanishing velocity. Altogether, both curves exhibit a more complex shape compared to the nearly linear behavior observed in earlier studies~\cite{liao2020dynamical, redner2013structure, stenhammar2013continuum, stenhammar2014phase}.  However, already in Fig.~\ref{fig:vel-rho}, the linear behavior of $v(\rho)$ is evident at small densities, as the contribution of those particles in dense areas  is insignificant at these densities. Partially, this can be explained by our choice of $\epsilon$ and $D_T$, see Appendix. 

 An important aspect in the present context is that the decrease of $v$ with $\rho$ is much less prominent in presence of fluctuations, particularly in the low-density regime. This confirms our picture that the trapping effect in the modified ABP model in hindered. 
 
In fact, using the function  $v(\rho)$  one can  obtain, in the sprit of earlier works on conventional ABPs~\cite{tailleur2008statistical}, an 
 effective free energy of the system as a function the density $f(\rho)$. This function  is composed of two contributions, the bulk contribution, $f_0$, and a contribution due to the repulsive interaction of the particles , $f_\mathrm{rep}$, yielding 
   \begin{equation}                                                                                                                       
     f(\rho) = f_0(\rho) + f_\mathrm{rep}(\rho)~,
\end{equation}
with
\begin{equation} \label{eq:}
  f_0(\rho) = \rho (\text{ln}\rho - 1 ) + \int_{0}^{\rho} \text{ln} [v(\rho)] d \rho
\end{equation}
and
\begin{equation}
  f_{rep}(\rho) = k_\mathrm{rep} \Theta(\rho - \rho_t)(\rho - \rho_t)^4~.
\label{rep}
\end{equation}

\begin{figure}[ht]
  \centering
  \includegraphics[width=0.45\textwidth]{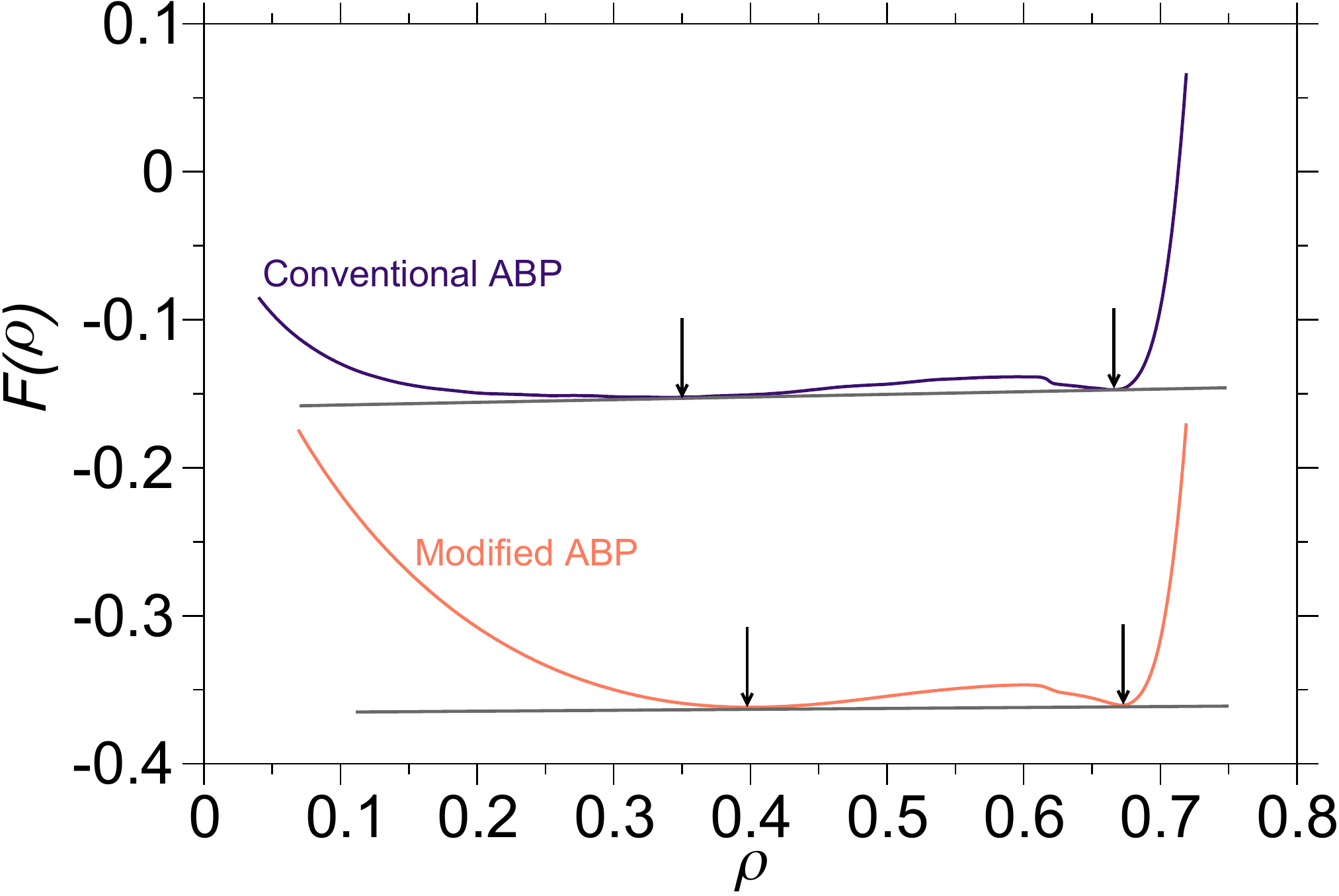}
  \caption{ The effective free energy as a function of the density for a modified ABP system with $D^\star_R=0.1$, $D^\star_T=0.001$, $v^\star=10$, \modified{$n=1$}, and the corresponding system without fluctuations. The dashed lines represent  the common tangent  line. One should note that for clarity the free energies are shifted by adding linear terms in $\rho$. These shifts are irrelevant for the common tangent construction.\label{fig:free-energy} }
\end{figure}

In Eq.~(\ref{rep}) $\Theta$ is the Heaviside function, $k_\mathrm{rep}$ is a  constant, and $\rho_t$ is a threshold density. Both of these parameters depend on the systems considered. Here, we set $k_\mathrm{rep}=10000$  and $\rho_t = 0.65$. The value of $k_\mathrm{rep}$ is chosen such that it is larger than the values chosen in Ref.~\cite{liao2020dynamical},  as the potential used here is closer to hard-core interaction. Nevertheless, it has the same order of magnitude as the values in Ref.~\cite{liao2020dynamical}. The choice of $\rho_t$ is guided by the largest local density values occurring in the system (see Fig.~\ref{fig:MPIS-Y1-no-fluctuation}). Using these parameters, we have calculated the free energy as a function of density, see Fig.~\ref{fig:free-energy}.  From these functions, we  obtain the coexistence densities via the common tangent construction. The resulting  coexistence densities show that the density associated with the dense phase is not affected by the fluctuations. In contrast, the density of the dilute phase is shifted to larger values due to the fluctuations. This is in line with the numerical  findings  in Fig.~\ref{fig:MPIS-Y1-no-fluctuation}.

\section{Conclusions}
In the present paper, we studied the impact of  fluctuations on the dynamics of  single ABPs,  as well as their collective behavior. Fluctuations are introduced to model a dynamically  evolving environment in which the particle diffuses  and  explores regions with different viscosities. This corresponds  essentially to the idea of diffusing diffusivity model. More specifically, we assumed that the motility, rotational and the translational diffusivities fluctuate simultaneously according to a random process that mimics the heterogeneity of the system. Using the sub-ordination concept, we obtained the exact form of the MSD and the asymptotic behavior of the PDF  of particle displacement for a general random process. We continued by calculating  explicit forms of the MSD and the PDF of particle displacements, assuming a square Ornstein-Uhlenbeck  process.  We considered two different initial conditions of the fluctuating process, namely  equilibrium and the non-equilibrium initial conditions. We showed that, similar to the case of passive particles, the PDF of particle displacements exhibits an exponential behaviour at short times compared to the characteristic time of the fluctuation. At longer times, the Gaussian form of the distribution is recovered. The MSD of the fluctuating ABP preserves the  ballistic and linear regimes familiar from  the ordinary ABP. However,  complicated cross-over effects appear when the characteristic time for the fluctuations is much shorter than the rotational relaxation time. Our analytical results for a single particle are fully confirmed by numerical calculations. 

 The second part of our study was devoted to the impact of the fluctuations on the collective behavior of the system, using particle based simulations. We focused on the motility induced phase separation phenomenon, which is well established  for ordinary ABPs. To this end, we calculated the phase diagram in the   motility-density  plane and compared it with that for the ordinary ABP  system. The structure of the formed clusters contain more roughness on the boundaries as well as disorders in the body of the formed clusters in the modified ABP system. Using our theoretical results, we argued that these observations in the collective system are mainly due to the short time behavior of the PDF of the translational and the rotational displacements. To support our discussion, we studied the density dependent of the particle velocity. We demonstrated  that, a  trapping effect of  particles due to collisions becomes less effective when  fluctuations are introduced to the system. 

Our main result in studying the collective behavior of the system, namely the hindrance of the MIPS, displays  a similarity to the effect of  environmental disorders on the collective dynamics of ordinary ABPs~\cite{ro2021disorder}. Such an emerging similarity, thus,  suggests that  environmental disorders can, to some extent,  be modeled  by including  fluctuations in the equations of motion  of ABPs.  In our model,  different features of the fluctuations can be manually controlled by adjusting the  characteristic parameters of the  sub-ordinator process, such as noise intensity, correlation time and  its dimension. Therefore,  a quantitative study on the impact of these parameters on the  collective behavior of ABPs on the one hand, and that of different densities or types of disorders on the other hand, will be interesting.



\bibliography{main-final.bbl} 
\bibliographystyle{apsrev4-2} 

\appendix
\section{Effect of potential hardness and the translational diffusion constant on the function $v(\rho)$}\label{append1}
In this Appendix we  explore in more detail the  dependency  of the velocity on the local density, where  Fig.~\ref{fig:vel-rho} reveals the commonly described  behavior that is, a decrease of $v$ with $\rho$ . It is,  however, not linear in $\rho$, thus  deviating from  the  behavior observed in~\cite{liao2020dynamical, redner2013structure, stenhammar2013continuum, stenhammar2014phase}.   The main reason behind this discrepancy is the fact that, in calculating $v(\rho)$, we  average over different regions in space where the MIPS has already occurred. In other words, we consider both the dilute and dense areas. In contrast,  in other studies one considers only the dilute regime. 

   To support our argument, we repeat our calculation by simulating a situation, where the MIPS is significantly hindered.
In order to manually prevent MIPS, one can  reduce the hardness of the particles by modifying the parameter $\epsilon$ and/or increasing the translational diffusion coefficient, $D^\star_T$. In both cases, the trapping effect becomes less pronounced  and thus, the MIPS is hindered.  We present the results for $v(\rho)$  by considering different values of  $\epsilon$  and  $D^\star_T$ in Fig.~9.

\begin{figure}[ht]
  \centering
  \includegraphics[width=0.49\textwidth]{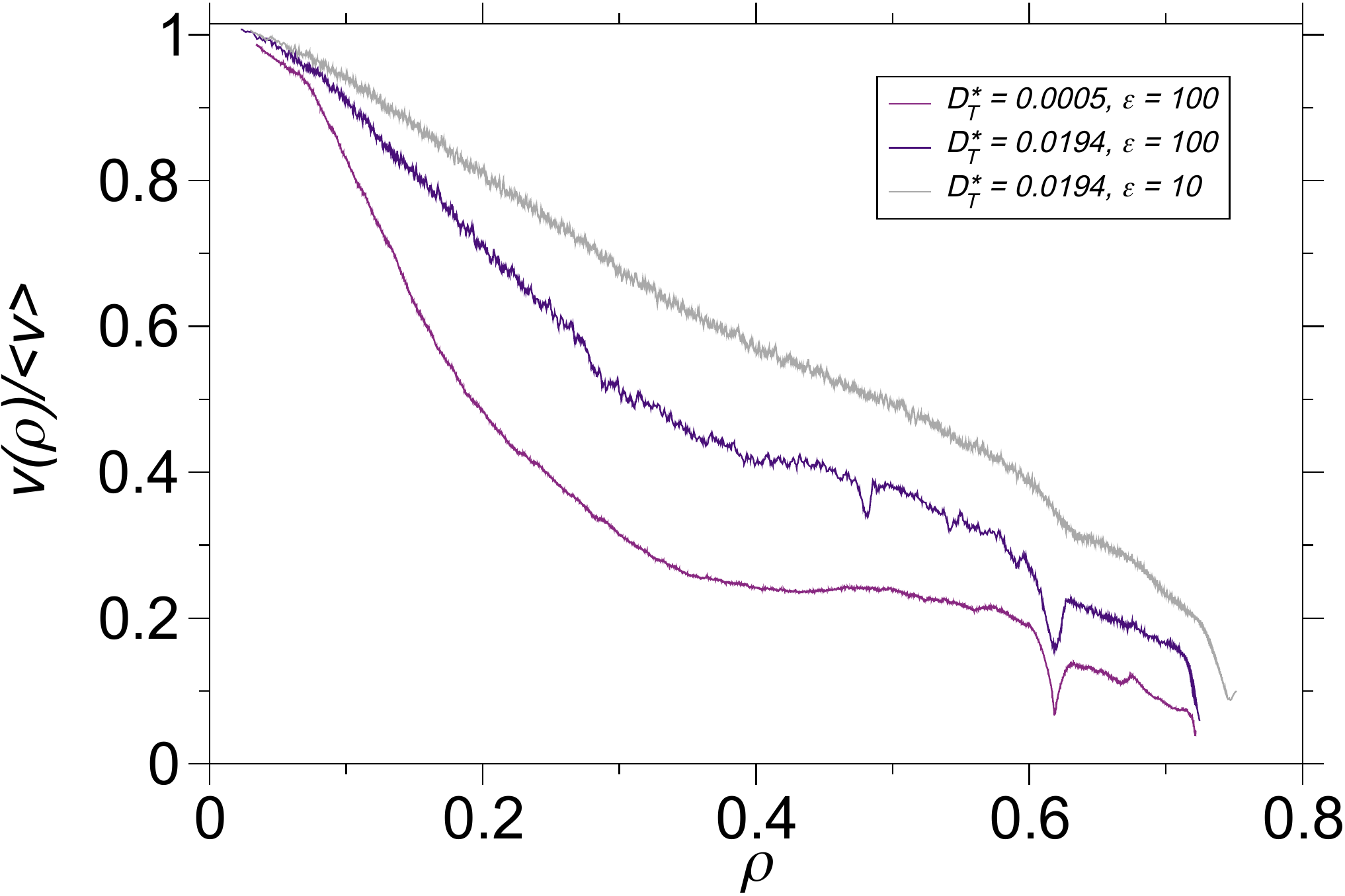}
  \caption{\label{fig:comparison}\nhs{The dependence of the local velocity on the local density for  conventional ABP systems. All of the considered systems have  $D_R=0.05$,  $v_0=10$, and $\bar{\Phi}=0.44$~.  These systems differ from each other in their values of $D_T$ and the hardness of their inter-particular interactions.}}
\end{figure}

The bottom curve in Fig.~9  corresponds to a very small value of $D^\star_T=0.0005$ together with a large  $\epsilon=100$. These values are associated with a strong occurrence of MIPS, yielding an extreme non-linearity, even  a  non-monotonicity   of  $v(\rho)$  at high densities.   By increasing the translational  diffusion coefficient to   $D^\star_T=0.0194$  and keeping $\epsilon=100$, a tendency towards a linear behaviour is observed in the middle curve. In fact, more particles are found in the dilute regime as they can easily scape from the transiently formed  clusters. Finally, making  particles softer by reducing $\epsilon$ yields   a bulk system  of a uniform density   where the $v-\rho$  curve becomes almost linear. In this case, averaging over different regions in space, does not lead to a discrepancy between our result with that in Refs.~\cite{liao2020dynamical, redner2013structure, stenhammar2013continuum, stenhammar2014phase}, as  no persisting dense regime is present. We also observe that  $v(\rho)$  decays almost linearly at small densities in all cases, as expected.


\end{document}